\documentclass[useAMS,usenatbib]{mn2e} 
\usepackage{graphicx,epsfig} 
\usepackage{epsf}
\usepackage{aas_macros}
\usepackage{appendix}
\usepackage[usenames]{color}

\newcommand{\ion}[2]{#1\,{\small #2}}

\begin{document}

\title[Pulsational Amplitude Growth]{Pulsational amplitude growth of the star KIC\,3429637 (HD\,178875) in the context of Am and $\rho$\,Pup stars}

\author[S.J. Murphy et al.] 
{Simon J. Murphy$^{1}$\thanks{email: smurphy6@uclan.ac.uk}, A. Grigahc\`ene$^{2, 3}$, E. Niemczura$^{4}$, D.W. Kurtz$^{1}$, 
\newauthor{K. Uytterhoeven$^{5,6}$}\\
\\
$^{1}$Jeremiah Horrocks Institute, University of Central Lancashire, Preston PR1 2HE\\
$^{2}$Centro de Astrof\'isica, Faculdade de Ci\^encias, Universidade do Porto, Rua das Estrelas, 4150-762 Porto, Portugal\\
$^{3}$Kavli Institute for Theoretical Physics and Department of Physics Kohn Hall, University of California, Santa Barbara, CA 93106, USA\\
$^{4}$Astronomical Institute, Wroc\l{}aw University, Kopernika 11, 51-622 Wroc\l{}aw, Poland\\
$^{5}$Instituto de Astrof\'{\i}sica de Canarias (IAC), Calle Via Lactea s/n, 38205 La Laguna, Tenerife, Spain\\
$^{6}$Dept. Astrof\'{\i}sica, Universidad de La Laguna (ULL), Tenerife, Spain\\}

\maketitle 

\begin{abstract}
KIC\,3429637 (HD\,178875) is a $\delta$\,Sct star whose light-curve shows continuous pulsational amplitude growth in \textit{Kepler} Mission photometry. Analysis of the three largest amplitude peaks in the Fourier transform indicates different growth rates for all three. We have ruled out instrumental causes, and determine the amplitude growth to be intrinsic to the star. We calculate time-dependent convection models and compare them with the observations. We confirm earlier characterisations that KIC\,3429637 is a marginal Am star through the analysis of new spectroscopic data. With the data presently available, a plausible cause of the amplitude growth is increasing pulsational driving as evolutionary changes shift the \ion{He}{II} driving zone deeper in this $\rho$\,Puppis star. If this model is correct, then we are watching real-time stellar evolutionary changes.
\end{abstract}


\section{Introduction}

The $\delta$\,Sct stars are low-overtone pressure mode pulsators on or near the main sequence between spectral types A2V to F0V and A3III to F5III \citep{kurtz2000}, whose pulsation frequencies range from 3 to 80\,d$^{-1}$ (i.e., 18\,min to 8\,h in period). Their pulsations are driven by the opacity ($\kappa$) mechanism operating on the \ion{He}{II} partial ionisation zone. These intermediate mass ($1.5-2.5$\,M$_{\sun}$) stars are mostly Pop.\,I stars, but some Pop.\,II $\delta$\,Sct stars do exist and are known as SX\,Phe variables \citep[see, e.g.][]{templetonetal2002}.

The $\delta$\,Sct instability strip is located at the junction between the classical Cepheid instability strip and the main sequence. It is not populated solely with $\delta$\,Sct variables with normal spectra, however. The $\delta$\,Sct stars have as their neighbours: the photometrically less variable, chemically peculiar, classical metallic-lined (Am) stars; the marginal Am (Am:, pronounced ``Am colon'') stars with slightly milder abundance anomalies; the early or hot Am stars; the $\rho$\,Pup stars (evolved Am stars); the Ap stars (where the `p' means peculiar); the $\gamma$\,Dor stars; the $\lambda$\,Boo stars; and the `normal' A-stars. This paper concerns KIC\,3429637, a $\rho$\,Pup star showing $\delta$\,Sct pulsations.

\subsection{The Am stars}

The classical Am stars are those whose \ion{Ca}{II} K-line types appear too early for their hydrogen line types, and metallic-line types appear too late, such that the spectral types inferred from the \ion{Ca}{II} K- and metal-lines differ by five or more spectral subclasses. Such stars are therefore commonly given three spectral types, e.g. Am\,kA8hA9mF3, corresponding to the K-line, hydrogen-lines and metallic-lines, respectively. The Am group makes up a significant fraction of late A stars -- up to 50\,per\,cent at A8 (\citealt{smith1973}; \citealt{smalleyetal2011}), and features abundance anomalies typically of factors of $\pm$10 \citep{abt2009}. Historically, the classical Am stars were thought not to pulsate, but with the micromagnitude photometric precision of the space-based \textit{Kepler} mission six of the ten known Am stars in the \textit{Kepler} field of view are observed to pulsate \citep{balonaetal2011}, and \citet{smalleyetal2011} found 200 of 1600 metallic-lined A and F dwarfs to be pulsating from SuperWASP data.

The Am stars rotate sufficiently slowly that turbulent motions do not prevent gravitational settling of helium \citep{baglinetal1973}. When enough helium is drained from the \ion{He}{II} partial ionisation zone, the star no longer pulsates. Meanwhile, ions with absorption lines near the peak wavelength of the photon energy distribution are radiatively levitated towards the surface, and others, notably He, C, Ca and Sc, gravitationally settle, thus providing the abundance anomalies \citep{kurtz1989}. 

The marginal Am stars show fewer than five spectral subclasses between the \ion{Ca}{II} K and metallic lines. Such stars are therefore less extreme examples of the classical Am stars, with milder abundance anomalies. \citet{kurtz1978} showed the majority of the Am: stars lie at the blue edge of the $\delta$\,Sct instability strip (his Fig.\,2), which makes them difficult to distinguish from the `hot Am stars' whose primary characteristics are Am anomalies in stars hotter than A4. Marginal Am stars are also found at the cool edge of the instability strip, however, with spectra similar to the historical $\delta$\,Del stars. We refer the reader to \citet{gray&garrison1989} for a discussion of why the $\delta$\,Del classification has been dropped and why the evolved Am stars are now known as $\rho$\,Pup stars.

\citet{kurtz1978} presented members of the $\rho$\,Pup class that also pulsate. A particularly good case-study of a $\rho$\,Pup star that is also a high-amplitude $\delta$\,Sct (HADS) star is that of HD\,40765 \citep{kurtzetal1995}. There it was found that spectral peculiarities still remain in the presence of a peak-to-peak surface radial velocity range of 14\,km\,s$^{-1}$.

\subsection{Models of pulsating Am stars}

The introduction of time-dependent convection in the modelling of $\delta$\,Sct star pulsations was a major step towards a quantitative comparison between theory and observations. \citeauthor{dupretetal2004} (\citeyear{dupretetal2004}, \citeyear{dupretetal2005}) have successfully explained the red edge of the $\delta$\,Sct instability strip by using these models. This approach was successfully applied to three $\delta$ Scuti stars by \citet{dupretetal2005}. The $\delta$\,Sct stars observed by \textit{Kepler} offer a higher-level test for the theory. In this work we apply time-dependent convection models to a $\delta$\,Sct star observed by \textit{Kepler} for the first time.

Before we move on to the specific case of KIC\,3429637, let us discuss the issue of pulsation coincident with metallicism, i.e. of the $\rho$\,Pup stars. In the Am stars it was thought that there should be insufficient helium left in the \ion{He}{II} partial ionisation zone for $\delta$\,Sct pulsations to be driven, but this is becoming increasingly doubted (e.g. \citealt{catanzaro&balona2012}). \citet{kurtz1989} discussed the theoretical exclusion of metallicism and pulsation in detail before announcing the first case of an extreme classical Am star showing $\delta$\,Sct pulsations. It appears that pulsation in chemically peculiar stars with pulsation velocities of several hundred m\,s$^{-1}$ can be so laminar that turbulence on the order of cm\,s$^{-1}$ is not generated. If turbulence were generated, the diffusion process would become ineffective and the stars would become homogenised. \citet{turcotteetal2000} addressed the issue, noting that it is velocity gradients that generate turbulence, and not the speed of displacement itself -- fast but uniform displacement will not become turbulent. Their models, the New Montreal Models, which take diffusion of elements up to Ni into account, find stable stars with He abundances as low as 0.114 at 750\,Myr. Specifically, He is still substantially present in the \ion{He}{II} driving region in these models, and the iron-peak-element opacity bump is increased relative to standard models and contributes to excitation of longer-period modes. As the stars evolve, they begin to pulsate naturally (i.e. without invoking dredge-up or hypothetical mass-loss \citep{turcotteetal2000}) because of the way that radius expansion causes the \ion{He}{II} ionisation zone to shift inward in mass fraction where more residual He is present \citep{coxetal1979}. The evolution leads to the increase in period of observed pulsation modes, and \citeauthor{turcotteetal2000} concluded that generally the effect of diffusion is to stabilise against p-mode pulsations but excite g-modes. No extensive observational study covering a large sample of stars has been carried out to investigate pulsation frequency ranges in light of abundance anomalies.

In spite of all the strengths of diffusion theory (see e.g. \citealt{kurtz2000}), the diffusion models still fail to reproduce the recent observations of the commonality of pulsating Am stars \citep{smalleyetal2011} satisfactorily, though we note that the disparity is not as great as it seems because 24\,per\,cent of their ``pulsating Am'' sample were actually ``Fm $\delta$\,Del" stars, and that pulsations are expected to occur naturally in these evolved stars according to modern diffusion models. We refer the reader to the work of \citet{balonaetal2011a} for further discussion and observations on the edges of the instability strip for pulsating Am stars.

\subsection{The Kepler Space Mission}

The Kepler Space Mission features an array of 42 CCDs covering a 115\,deg$^2$ area of the sky between the constellations of Cygnus and Lyra. It captures white-light photometric data from $\sim$150\,000 stars, with the objective of detecting transiting planets that orbit within the habitable zone \citep{kochetal2010}. The attained precision is around the micromagnitude level, and, with a duty cycle $>$\,92\,per\,cent, the mission offers vast advances in astrophysics through the study of stellar pulsations. With photometry alone, transiting studies only yield the ratio of the planet and host-star radii, but asteroseismology can determine stellar radii to within $\sim$\,1\,per\,cent in some cases \citep{gillilandetal2010a}. With that in mind, about 1\,per\,cent of observations are allocated to asteroseismology.

\textit{Kepler} data are available in two cadences: long-cadence (LC) with effective 29.4-min integrations; and short-cadence (SC) with 58.9-s integrations. The near-continuous observations are interrupted every 32\,d (one `month') for data downlinking, and every three months corresponds to one quarter of \textit{Kepler}'s 372.5-d heliocentric, Earth-trailing orbit. LC data are therefore available in quarters (denoted Qn), and SC data are further subdivided into months (denoted Qn.m). Each quarter the satellite performs a roll to keep its solar panels pointed towards the Sun and its radiator pointed towards deep space. Stars subsequently fall on to a different detector module when the symmetrical focal plane rotates. The unfortunate failure of module 3 early in the mission means that a small fraction ($\sim \frac{4}{21}$) of the field of view is only observable for three quarters of the year.

\section{Photometric Observations}
\label{sec:photometry}

\begin{figure*}
\begin{center}
\includegraphics[width=0.9\textwidth,natwidth=720,natheight=360]{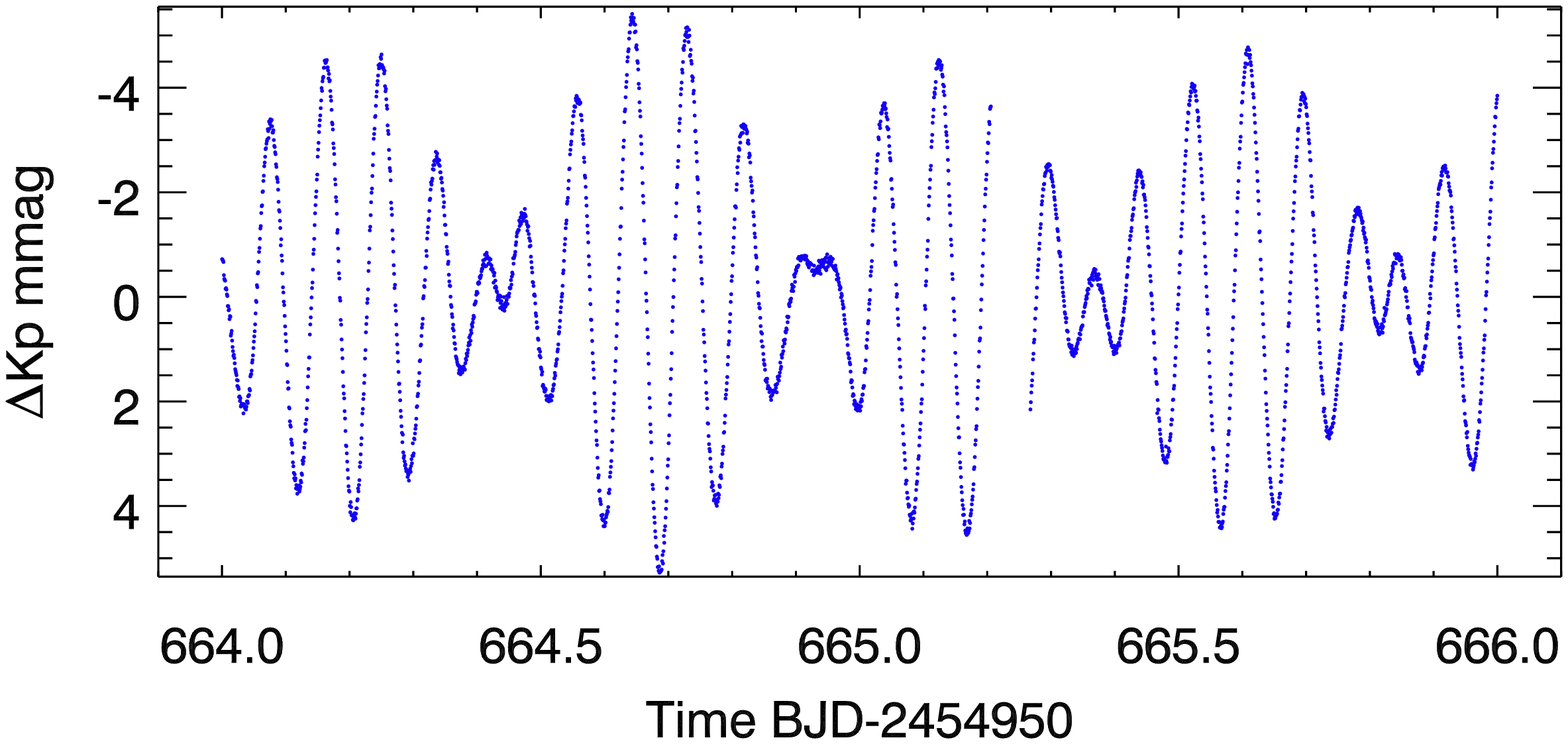}
\includegraphics[width=0.9\textwidth,natwidth=720,natheight=360]{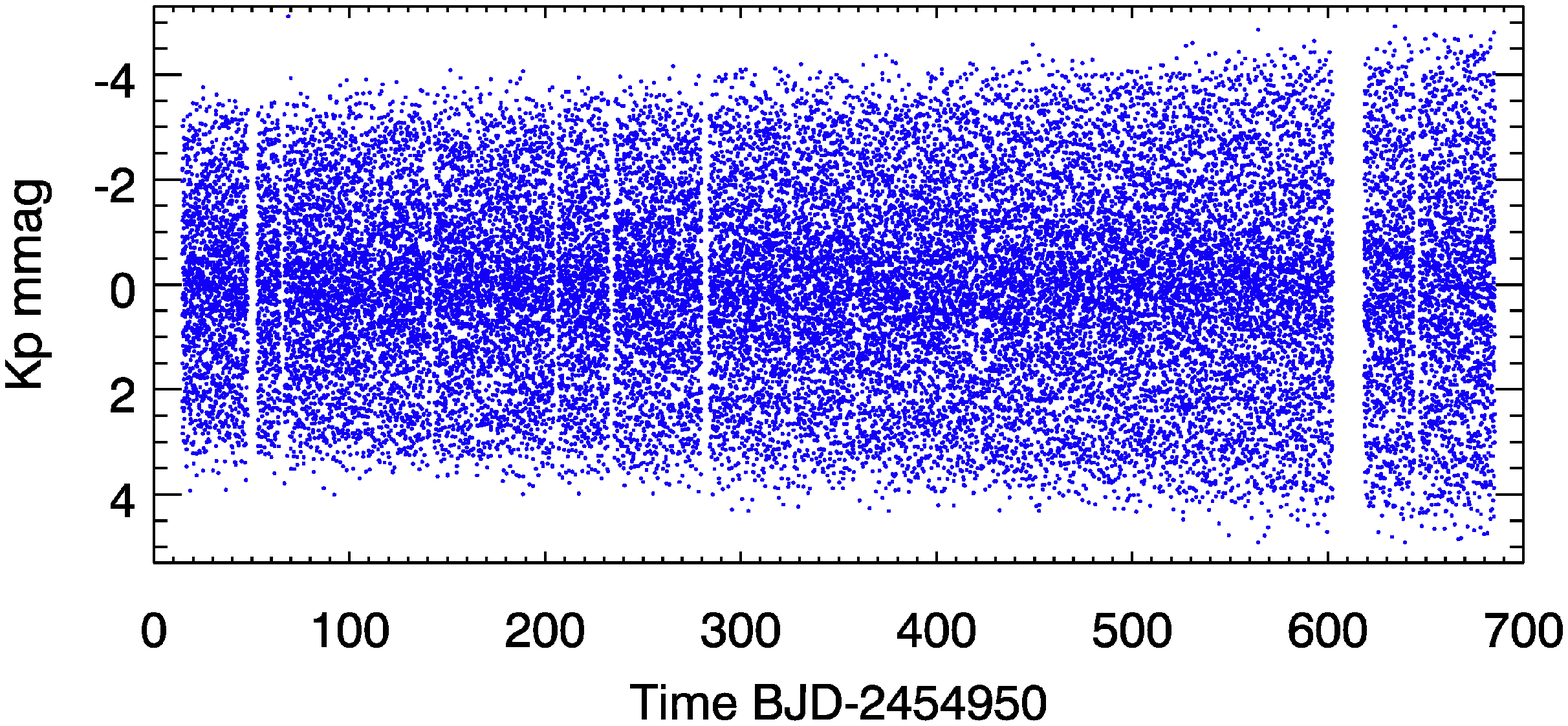}
\caption{Upper Panel: the $\delta$\,Sct pulsations of KIC\,3429637 in SC PDC from Q8. Kp denotes \textit{Kepler} magnitude, and BJD is Barycentric Julian Date. Lower Panel: the unedited light-curve of the LC PDC data from Q1 to Q8. The amplitude of the light variation is clearly growing with time, as can be seen from the unresolved envelope of the pulsations.}
\label{fig:lightcurve}
\end{center}
\end{figure*}

\textit{Kepler} data are available in abundance for KIC\,3429637, which is the 9th brightest in the $\delta$\,Sct working group of the Kepler Asteroseismic Science Consortium (KASC). Long-cadence (LC) public data are available from Q1 through to Q8, and short-cadence (SC) public data are available for each month of Q7 and Q8. The LC data, with just over 30\,000 data points covering 670.33\,d, have a duty cycle of 91.4\,per\,cent. The light-curve for the Pre-search Data Conditioned (PDC) LC data is displayed in Fig.\,\ref{fig:lightcurve}, from which it is immediately clear that the amplitude of the light variation is growing with time. We used only the Least-Squares version of the PDC pipeline (i.e. PDC-LS).

\begin{figure*}
\begin{center}
\includegraphics[width=0.9\textwidth]{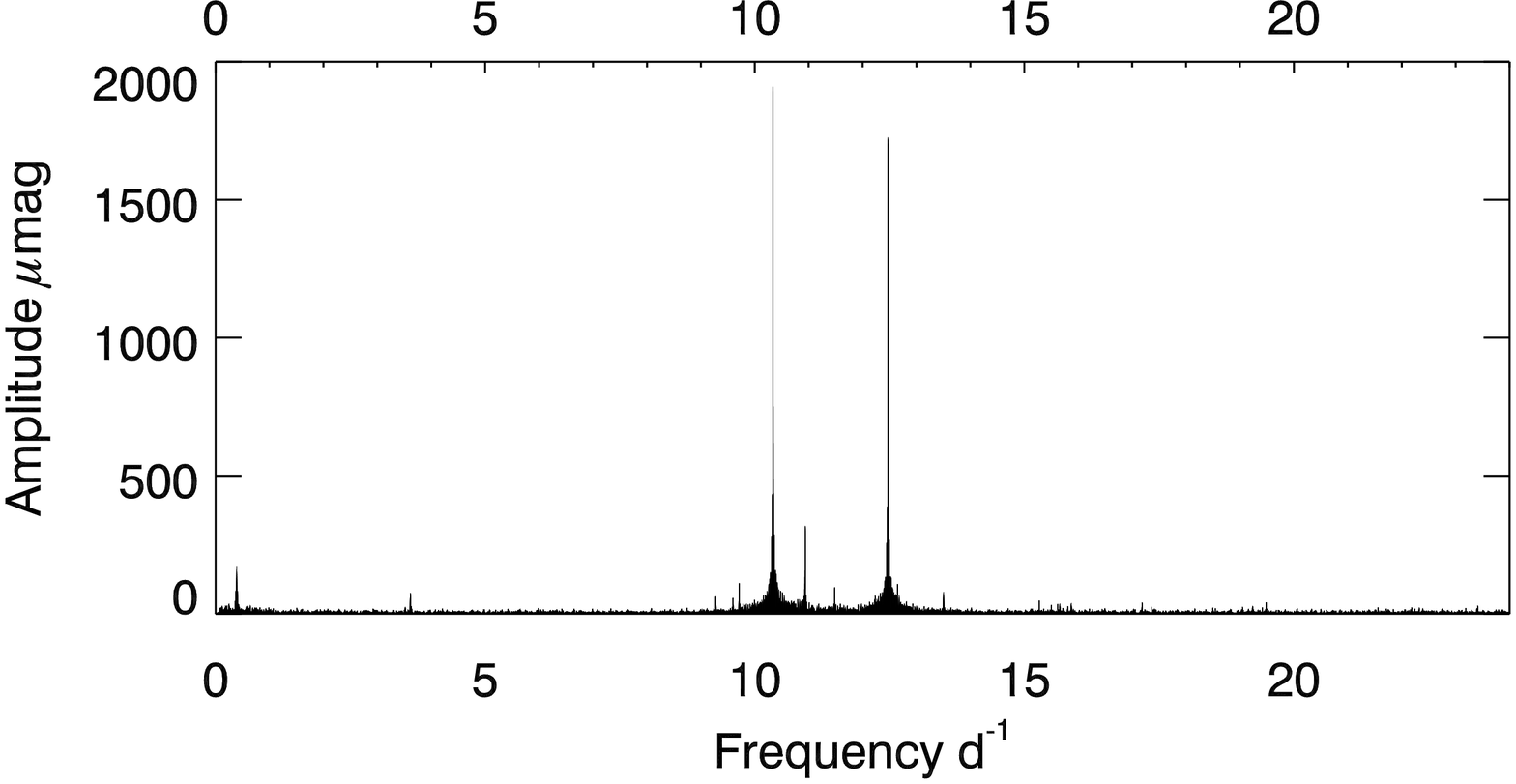}
\caption{The Fourier transform of the unedited Q5 LC time-string. The three highest-amplitude peaks, in descending amplitude order, are $f_1$, $f_2$ and $f_3$.}
\label{fig:fourier_0-24}
\end{center}
\end{figure*}

There are two obvious potential sources of the apparent amplitude growth: (1) the amplitude growth is intrinsic to the star, i.e. astrophysical; and (2) the amplitude growth is in some way instrumental. PDC data are not created to facilitate asteroseismology, but rather to prepare the data for planet searches. As such, the data are not to be used for asteroseismology without caution that the pipeline might modify stellar variability. To test whether the pipeline caused the amplitude growth we compared the PDC data to the simple aperture photometry (SAP) data, for which only basic calibration has been performed and which is deemed suitable for asteroseismology, provided that instrumental trends are corrected for. Since instrumental trends generally affect low frequencies only, these are not an issue in the periodogram at the frequencies of pulsation shown in the Fourier transform of the data in Fig.\,\ref{fig:fourier_0-24}. The amplitudes of the pulsations in the PDC data were compared to the uncorrected SAP data, by means of a non-linear least-squares fit of the frequencies to each quarter of data using the software package {\small PERIOD04} \citep{lenz&breger2004}. The amplitude growth was the same in both data sets.

\begin{figure*}
\begin{center}
\includegraphics[width=0.9\textwidth]{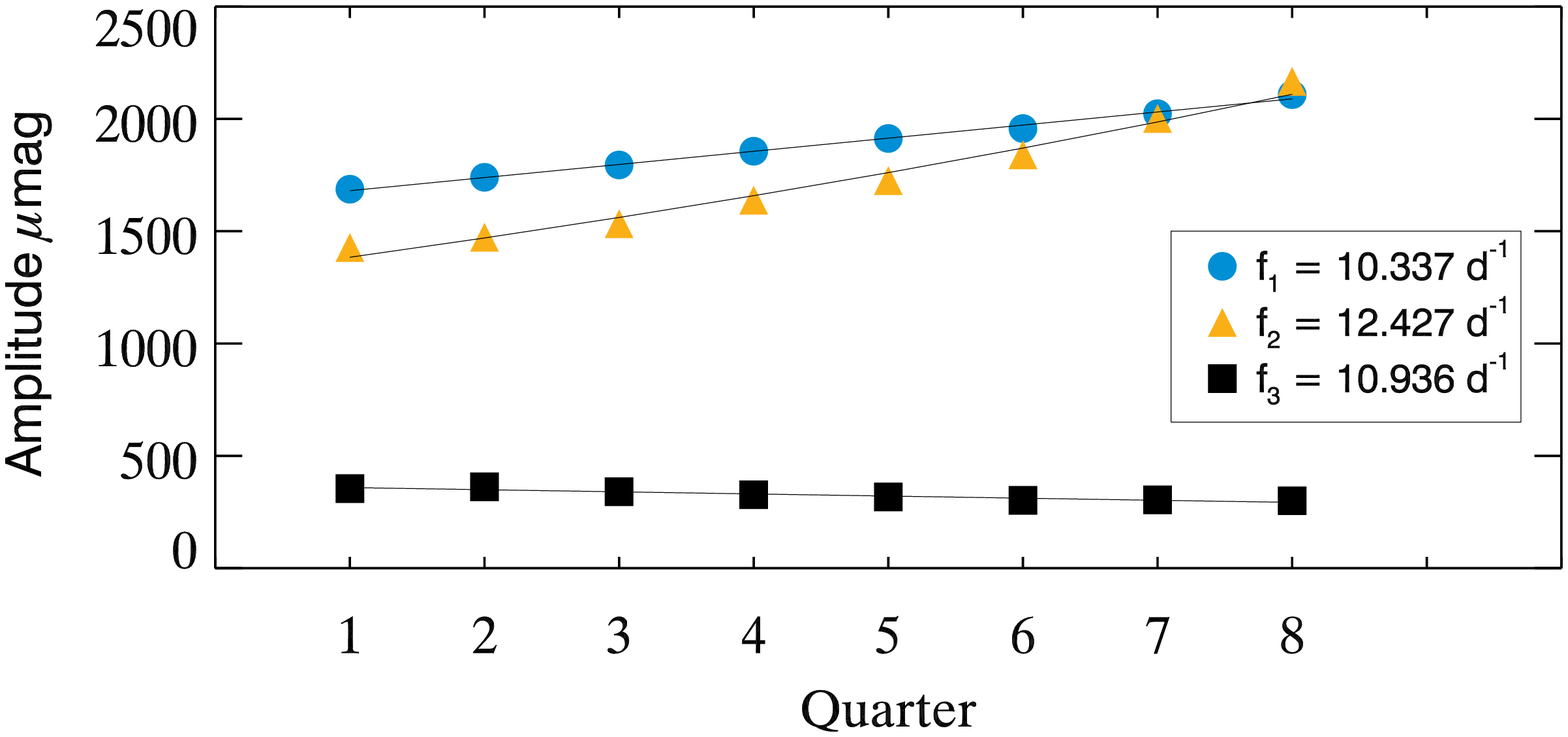}
\caption{The growth of the amplitude of $f_1$ and $f_2$, and decrease in amplitude of $f_3$ in LC PDC-LS flux as a function of time. Linear fits are shown for $f_1$ and $f_3$, and an exponential fit is shown for $f_2$. Errors on amplitude are on the order of a few micromagnitudes, and are much smaller than the plot symbols.}
\label{fig:growth}
\end{center}
\end{figure*}

The star has two dominant frequencies ($f_1 = 10.337$\,d$^{-1}$ and $f_2 = 12.472$\,d$^{-1}$), with amplitudes growing between 1.5 and 2.5\,mmag, and a third mode that lies between them ($f_3 = 10.936$\,d$^{-1}$) with an amplitude of $\sim$0.3\,mmag -- observable without any prewhitening. These frequencies are typical for $\delta$\,Sct stars. As an additional check that the amplitude growth is astrophysical and not instrumental, the amplitude of that third peak was tracked across multiple quarters for comparison. It was found that although the two main peaks continued to grow in amplitude across all quarters, $f_3$ showed a continual decrease in amplitude, as seen in Fig.\,\ref{fig:growth}. \textit{This proves that the growth in amplitude of the light variation is astrophysical}. In addition to these three main frequencies, some $\sim$40 statistically significant peaks with lower amplitudes exist in the data. We tracked the amplitudes of only the next four highest-amplitude modes, and have hence tracked all modes for which modelling was performed (see \S\,6). The amplitudes of these four modes, not shown in Fig.\,\ref{fig:growth}, show changes of no more than 30\,$\mu$mag across the 8 quarters and the changes appear to be noise-dominated, so the amplitudes of peaks with lower statistical significance were not tracked.

The optimum aperture for a star is defined such that the signal-to-noise ratio of the light captured from that star is maximised. This doesn't, therefore, capture all light from that star, because pixels further from the centre of the star contain less of the star's light and therefore relatively higher noise levels. It is possible, as a direct consequence of the choice of optimum aperture, that slightly more light from this star is being captured in each quarter and slightly less of a neighbouring star that is contributing towards the total light, which would create the effect of amplitude growth of \textit{all} pulsation frequencies in the star's Fourier spectrum. Although not all modes are growing in amplitude, a check is necessary. The Kepler Input Catalogue (KIC; \citealt{lathametal2005}; \citealt{brownetal2011}) lists the contamination for this star to be 0.037, meaning that about 3.7\,per\,cent of the light in the aperture of this star is coming from one or more neighbours rather than the target star itself. To rule out the hypothesis that the amplitude growth is due to changing flux fractions of any neighbour, we took the target pixel light curves of Q1 to Q5 and defined a mask for each one that covered every pixel available for the star for that quarter. We found that just as for the PDC LS pipeline data, with our custom mask the amplitudes of the dominant modes grew each quarter and the amplitude of the third mode that lies between them continued to decrease. 

In the CCDM (Catalogue of the Components of Double and Multiple stars; \citealt{donmagnet&nys1994}), KIC\,3429637 is listed as a double star with separation 1.58\,arcsec. The primary has $V = 7.8$ and the secondary has $V = 12.7$, thus the secondary has a spectral type around early K. Having a Hipparcos parallax of $3.75 \pm 0.58$\,mas, the stars are well separated with an orbital period $>1000$\,y, so tidal effects on the primary's rotation period are unimportant. Without radial velocities, we must treat KIC\,3429637 as if it were a single star. Furthermore, there is no evidence at present to rule out chance alignment.

We wish to examine the possibility that the amplitude change is caused by the secondary. According to the CCDM, the dimmer star contributes just 1\,per\,cent of the light coming from the pair. For this dimmer star to cause an amplitude change of $\sim$30\,per\,cent is impossible: one would have to remove it thirty times to exact such a change.

\section{Spectroscopic Observations}
\label{sec:spectroscopy}

Various methods exist for inspection of a star for chemical peculiarity, and definitions of peculiarity in the literature are complex. In his Am star review, \citet{conti1970} summarised the peculiarities as ``deficient Ca (Sc) or overabundant heavier elements, or both.'' Indeed, \citet{conti1965} used the line depth ratio of \ion{Sc}{II} $\lambda$4246.8\,/\,\ion{Sr}{II} $\lambda$4215.5 as an indicator of peculiarity and \citet{smith1970} went further and used another criterion \ion{Sc}{II} $\lambda$4320.7\,/\,\ion{Y}{II} $\lambda$4309.6, because of the potential for blending of the \ion{Sc}{II} line with the \ion{Fe}{I} line (at $\lambda$4247.3) in fast rotators; \citet{smith1970} noted that a substantial fraction of Am stars may be nearly normal in Sc and Ca. Modern abundance surveys are consistent with long-established characterisations: \citet{gebranetal2010} found that: ``All Am stars in the Hyades are deficient in C and O and overabundant in elements heavier than Fe but not all are deficient in calcium and/or scandium.''

Another method for detection of peculiarity is the Str\"omgren $\Delta\,m_1$ index. A-stars with negative $\Delta\,m_1$ indices are commonly Am stars, with more negative values indicating a more chemically peculiar star. Using $uvby$ parameters for KIC\,3429637 from \citet{olsen1983} and \citet{hauck&mermilliod1998} ($b-y = 0.189$, $m_{1} = 0.215$, $c_{1} = 0.849$), noting that both sources agree to that precision, the $\Delta\,m_1$ index for this star was calculated using Table 1 of \citet{crawford1979} for calibration as $\Delta\,m_1=-0.028$, which indicates an Am/Am: star. Our evaluation is consistent with that of \citet{abt1984}, who classified the star as Am(F2/A9/F3) -- a marginal Am star given that the metallic-line type and Ca II K-line type differ by fewer than five spectral subclasses. We note that although the Crawford $\Delta\,m_1$ calibration is for main sequence stars, it may also be applied to evolved stars.

Atmospheric parameters for KIC\,3429637 were calculated by \citet{catanzaroetal2011} as part of their spectroscopic survey of potential \textit{Kepler} asteroseismic targets in the $\delta$\,Sct and $\gamma$\,Dor instability strips. They estimated an equivalent spectral type of F0\,III (from fundamental parameters), $v\sin i$ = 50\,$\pm$\,5\,km\,s$^{-1}$, and used a variety of methods to determine $T_{\rm eff}$ and $\log g$. The weighted mean of their values for $T_{\rm eff}$ and $\log g$ are 7300\,$\pm$\,200\,K and 3.16\,$\pm$\,0.25 (cgs), respectively. \citeauthor{catanzaroetal2011} also computed masses for this star: a canonical value of 3.6$^{+0.7}_{-0.6}$\,M$_{\sun}$ and a non-canonical value (that is, with convective overshooting: $\lambda_{\rm{OV}}=0.2\,H_p$) of 3.2$^{+0.6}_{-0.5}$\,M$_{\sun}$. They estimated the luminosity to be 20\,$\pm$\,6\,L$_{\sun}$. Additional parameters from the KIC, roughly derived from broad-band photometry, are $R = 4.1$\,R$_{\sun}$ and [Fe/H] = 0.12. It is immediately apparent from the luminosity class and $\log g$ parameters that the star is a giant, and thus an evolved Am, or $\rho$\,Pup, star. A comparison of the $m_{1}$ value of this star with those studied by \citet[][his Fig.\,3]{kurtz1976} strongly implies this star is of the $\rho$\,Pup type. We return to this in Section\,\ref{sec:cause}.

\begin{table*}
\centering
\caption{The elemental abundances for KIC\,3429637 as compared to the solar values of \citet{asplundetal2009}. Columns 4 and 5 display the abundances of this star and the Sun, respectively, with their standard deviations. For those elements in column 4 whose abundances were determined from only one or two lines, an accurate uncertainty estimate cannot be made so we provide the mean uncertainty of our other abundances (0.12) instead. Column 6 is the difference between columns 4 and 5. Column 7 contains [M/H] for the normal A-stars from \citet{gebranetal2010} for the elements they analysed. In their paper, these abundances were relative to the solar values of \citet{grevesse&sauval1998}, but here we correct them to the latest and most precise abundances: \citet{asplundetal2009}. Column 8 contains the difference between the abundances of this star and a normal A-star. The final column is like column 8, but with a metallicity correction applied -- it represents the difference between KIC\,3429637 and a normal star of the same Fe/H (see in-text explanation, Section\,\ref{sec:spec-diss}).}
\label{abundances}
\begin{tabular}{ccclcrrrr}
\hline
element & Z & \# of lines & abundance & solar value & [M/H] \hspace{2mm} & A-stars & $\Delta$\,M & $\Delta$\,M$_{\rm corr}$\\
\hline
C	&$	6	$&$	6	$&$	8.03	\pm	0.16	$&$	8.43	\pm	0.05	$&$	-0.40	\pm	0.17	$&$	-0.06	$&$	-0.25	$&$	-0.49	$	\\
Na	&$	11	$&$	1	$&$	6.07	\pm	0.12	$&$	6.24	\pm	0.04	$&$	-0.17	\pm	0.13	$&$	0.29	$&$	-0.37	$&$	-0.04	$	\\
Mg	&$	12	$&$	6	$&$	7.53	\pm	0.15	$&$	7.60	\pm	0.04	$&$	-0.07	\pm	0.16	$&$	0.13	$&$	-0.22	$&$	0.07	$	\\
Si	&$	14	$&$	4	$&$	7.32	\pm	0.17	$&$	7.51	\pm	0.03	$&$	-0.19	\pm	0.17	$&$	0.39	$&$	-0.54	$&$	-0.08	$	\\
S	&$	16	$&$	1	$&$	7.67	\pm	0.12	$&$	7.12	\pm	0.03	$&$	0.55	\pm	0.12	$&$	-	$&$	-	$&$	-	$	\\
Ca	&$	20	$&$	11	$&$	6.34	\pm	0.09	$&$	6.34	\pm	0.04	$&$	0.00	\pm	0.10	$&$	0.04	$&$	-0.02	$&$	-0.03	$	\\
Sc	&$	21	$&$	11	$&$	2.97	\pm	0.10	$&$	3.15	\pm	0.04	$&$	-0.18	\pm	0.11	$&$	-0.25	$&$	0.09	$&$	-0.38	$	\\
Ti	&$	22	$&$	29	$&$	4.82	\pm	0.13	$&$	4.95	\pm	0.05	$&$	-0.13	\pm	0.14	$&$	0.15	$&$	-0.21	$&$	-0.11	$	\\
V	&$	23	$&$	2	$&$	4.43	\pm	0.12	$&$	3.93	\pm	0.08	$&$	0.50	\pm	0.14	$&$	-	$&$	-	$&$	-	$	\\
Cr	&$	24	$&$	25	$&$	5.50	\pm	0.20	$&$	5.64	\pm	0.04	$&$	-0.14	\pm	0.20	$&$	0.17	$&$	-0.28	$&$	-0.05	$	\\
Mn	&$	25	$&$	6	$&$	5.24	\pm	0.14	$&$	5.43	\pm	0.05	$&$	-0.19	\pm	0.15	$&$	0.13	$&$	-0.36	$&$	-0.16	$	\\
Fe	&$	26	$&$	80	$&$	7.35	\pm	0.11	$&$	7.50	\pm	0.04	$&$	-0.15	\pm	0.12	$&$	0.19	$&$	-0.34	$&$	0.00	$	\\
Ni	&$	28	$&$	17	$&$	6.43	\pm	0.12	$&$	6.22	\pm	0.04	$&$	0.21	\pm	0.13	$&$	0.36	$&$	-0.12	$&$	0.39	$	\\
Cu	&$	29	$&$	2	$&$	3.92	\pm	0.12	$&$	4.19	\pm	0.04	$&$	-0.27	\pm	0.13	$&$	-	$&$	-	$&$	-	$	\\
Zn	&$	30	$&$	2	$&$	4.37	\pm	0.12	$&$	4.56	\pm	0.05	$&$	-0.19	\pm	0.13	$&$	-	$&$	-	$&$	-	$	\\
Sr	&$	38	$&$	2	$&$	3.97	\pm	0.12	$&$	2.87	\pm	0.07	$&$	1.10	\pm	0.14	$&$	0.39	$&$	0.81	$&$	1.41	$	\\
Y	&$	39	$&$	3	$&$	2.94	\pm	0.01	$&$	2.21	\pm	0.05	$&$	0.73	\pm	0.05	$&$	0.57	$&$	0.19	$&$	1.01	$	\\
Zr	&$	40	$&$	3	$&$	2.98	\pm	0.03	$&$	2.58	\pm	0.04	$&$	0.40	\pm	0.05	$&$	0.62	$&$	-0.20	$&$	0.10	$	\\
Ba	&$	56	$&$	3	$&$	3.08	\pm	0.13	$&$	2.18	\pm	0.09	$&$	0.90	\pm	0.16	$&$	-	$&$	-	$&$	-	$	\\
La	&$	57	$&$	1	$&$	0.92	\pm	0.12	$&$	1.10	\pm	0.04	$&$	-0.18	\pm	0.13	$&$	-	$&$	-	$&$	-	$	\\
\hline
\end{tabular}
\end{table*}

\subsection{New spectroscopic observations}

A medium-resolution optical spectrum of the investigated object was obtained using the cross-dispersed, Fibre-fed \'{E}chelle Spectrograph (FIES) on the 2.5-m Nordic Optical Telescope, at Roque de los Muchachos, La Palma during the run of 2010 Aug 3-5 (proposal: 61-NOT7/10A). FIES offers an optical spectrum from 3700 to 7300\,{\AA}, with a spectral resolution of 46\,000 in a single exposure. The signal-to-noise of the spectrum is about 100 at 5500\,{\AA}.

The spectrum was reduced using standard procedures of FIEStool, which consists of bias subtraction, extraction of scattered light produced by the optical system, division by a normalised flat-field, and wavelength calibration. After reduction, the spectrum was normalised to the continuum by using {\small SPLAT}, the spectral analysis tool from the Starlink project \citep{draperetal2005}.

\subsection{Atmospheric parameters determination}

To perform an abundance analysis, one needs to determine an appropriate atmospheric model of the star, which requires the knowledge of its effective temperature $T_{\rm eff}$, surface gravity $\log g$ and metallicity. The necessary atmospheric models were computed with the line-blanketed LTE {\small ATLAS9} code \citep{kurucz1993a}, which treats line opacity with opacity distribution functions (ODFs). The synthetic spectra were computed with the {\small SYNTHE} code \citep{kurucz1993b} over the wavelength ranges 4000-5850 and 6100-6800\AA, excluding those segments contaminated with telluric lines. Both codes, {\small ATLAS9} and {\small SYNTHE} were ported under GNU Linux by \citet{sbordone2005} and are available online\footnote{wwwuser.oat.ts.astro.it/atmos/}. The stellar line identification and the abundance analysis were performed on the basis of the line list from \citet{castelli&hubrig2004}\footnote{http://wwwuser.oat.ts.astro.it/castelli/grids.html}. 

We derived $T_{\rm eff}$ for KIC\,3429637 using the sensitivity of hydrogen line wings to temperature, following the method proposed by \citet{veer-menneret&megessier1996}. The effective temperature was estimated by computing the {\small ATLAS9} model atmosphere which gives the best match between the observed H$\delta$, H$\gamma$, H$\beta$ and H$\alpha$ line profiles and those computed with {\small SYNTHE}. The same value of $T_{\rm eff}$ was obtained from the analysis of iron lines. In this method effective temperature, surface gravity and microturbulence ($\xi$) are determined by the comparison of the abundances obtained from neutral and ionised iron lines. The analysis is based on Fe lines because they are the most numerous in the stellar spectrum. In general, we require that the abundances measured from \ion{Fe}{I} and \ion{Fe}{II} lines yield the same result. The absorption lines of \ion{Fe}{I} depend mainly on $T_{\rm eff}$, $\xi$ and metallicity, and are practically independent of $\log g$, whereas the \ion{Fe}{II} lines are mostly sensitive to $\log g$. First, we adjust $\xi$ until we see no correlation between iron abundances and line intensity for the \ion{Fe}{I} lines. Second, $T_{\rm eff}$ is changed until we see no trend in the abundance versus excitation potential of the atomic level causing the \ion{Fe}{I} lines. $\xi$ and $T_{\rm eff}$ are not independent of each other. Then $\log g$ is obtained by fitting the \ion{Fe}{II} and \ion{Fe}{I} lines and by requiring the same abundances from both neutral and ionised lines. 

The abundances of chemical elements were determined by the spectrum synthesis method. Our analysis follows the methodology presented in \citet{niemczuraetal2009} and relies on an efficient spectral synthesis based on a least-squares optimisation algorithm. This method allows for the simultaneous determination of various parameters involved with stellar spectra and consists of the minimisation of the deviation between the theoretical flux distribution and the observed normalised one. The synthetic spectrum depends on the stellar parameters, such as $T_{\rm eff}$, $\log g$, $\xi$, $v\sin i$ and the relative abundances of the elements. Some of these parameters have similar influence on stellar spectra and have to be known before the determination of the abundances of chemical elements. $T_{\rm eff}$, $\log g$ and $\xi$ are the input parameters. All other aforementioned parameters can be determined simultaneously because they produce detectable and different spectral signatures. The $v\sin i$ values are determined by comparing the shapes of observed metal line profiles with the computed profiles, as shown by \citet{gray2005}. 

\begin{figure*}
\begin{center}
\includegraphics[width=0.9\textwidth]{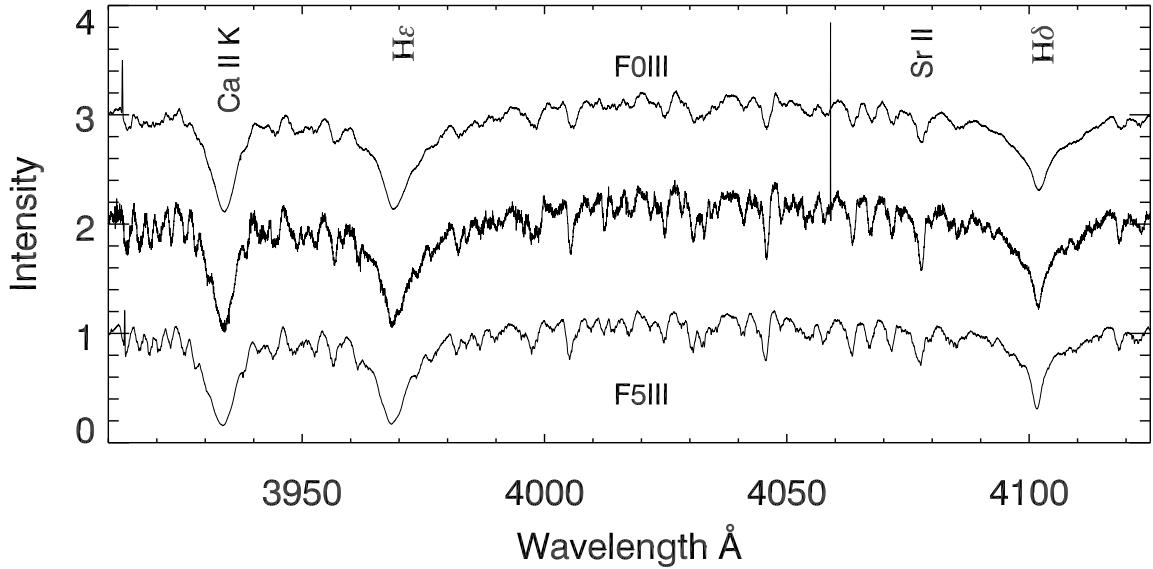}
\caption{Spectrum of KIC\,3429637 (centre) with two comparisons: the F0\,III star HD\,7312, and the F5\,III star HD\,65925, though neither is an MK-standard. \citet{catanzaroetal2011} listed KIC\,3429637 as F0\,III\,m. The hydrogen line type is a good match to $\sim$F0/A9. Despite being of the Am type, notice that the Ca\,II K-line is no weaker than the F0 comparison (cf. Table\,\ref{abundances}, in which calcium is completely normal). The metal lines are much stronger, though, and are even stronger than the F5 comparison. In particular, the \ion{Sr}{II} $\lambda$4077 line is much deeper, owing to the Am nature. The comparison spectra were obtained at ``http://www.eso.org/sci/observing/tools/uvespop/field\_stars\_uptonow.html'', normalised at 3910.0\,{\AA} and vertically separated by 1 unit.}
\label{fig:spectrum}
\end{center}
\end{figure*}

For the chemical abundance analysis we used short segments of the spectrum to isolate individual spectral features, and in a few cases, some blended features. In the case of blended lines, more than one chemical element can influence the line profile. Only the most important elements for characterising Am star peculiarities were considered. Every selected part of the spectrum was analysed by the spectrum synthesis method described above. We iteratively adjusted radial velocity, $v\sin i$, and chemical abundances until the determined parameters remained the same within 2\,per\,cent for three consecutive iterations, thus determining the closest match between the calculated and observed spectrum. Finally, we determined the average values of radial velocity, of $v\sin i$, and abundances of 20 elements with error estimates (Table\,\ref{abundances}). Table\,\ref{tab:atmos} summarises our determined atmospheric parameters, and we plot our spectrum against comparisons in Fig.\,\ref{fig:spectrum}.

\begin{table}
\centering
\caption{Fundamental atmospheric parameters determined from the FIES spectrum.}
\label{tab:atmos}
\begin{tabular}{cccc}
\hline
$T_{\rm eff}$ & $\log g$ & $\xi$ & $v\sin i$ \\
K & (cgs) & km\,s$^{-1}$ & km\,s$^{-1}$ \\
\hline
$7300 \pm 100$ & $3.0 \pm 0.1$ & $4.0 \pm 0.5$ & $51 \pm 1$ \\
\hline
\end{tabular}
\end{table}

\subsection{Discussion of spectroscopic results}
\label{sec:spec-diss}

The line ratios of \citet{conti1965} and \citet{smith1970} were inconclusive on the Am: nature of the star, necessitating the full abundance analysis (Table\,\ref{abundances}). We note normal Ca and nearly-normal Fe, but slightly deficient Sc, strongly deficient C, and highly overabundant Sr, Y and Ba that would collectively confirm that the star is chemically peculiar. However, observations of \citet{gebranetal2010} suggest that even normal A-stars are overabundant in Sr, Y and Zr with respect to solar values. This effect of overabundant heavy elements can have multiple origins:
\begin{enumerate}
\item[1.] The study of \citeauthor{gebranetal2010} focussed on the Hyades cluster, and those cluster stars could be systematically enriched in heavier elements. The abundances of nearly all heavier elements are known to be correlated with iron \citep{hill&landstreet1993}. The $\left[\frac{Fe}{H}\right]$ of the Hyades cluster is not well agreed upon. \citet{boesgaard&friel1990} determined a super-solar metallicity $\left<\left[\frac{Fe}{H}\right]\right> = 0.127\pm0.022$\,dex, yet \citet{varenne&monier1999} found a much lower value of $-0.05\pm0.03$\,dex when investigating 29 F-dwarfs. \citeauthor{gebranetal2010} themselves find $\frac{Fe}{H} = 0.19$\,dex above solar. A recent independent calculation \citep{carrera&pancino2011} based on three K-giants yields $\left<\left[\frac{Fe}{H}\right]\right> = +0.11\pm0.01$\,dex and is close to the mean of previous studies collated by those authors. We may therefore safely conclude at least a slightly super-solar Fe abundance and that correlation with Fe is a contributor.
\item[2.] Using the Sun as a standard star has the weakness of assuming the Sun is typical. \citet{melendezetal2009} found the Sun to be depleted in refractory elements by $\sim$20\,per\,cent relative to volatile elements compared to solar twins. Specifically, the lighter, volatile elements (e.g. C, N and O) are enriched by $\sim$0.05\,dex and the heavier, refractory elements depleted by $\sim$0.03\,dex. Their findings were confirmed in another study \citep{ramierezetal2009}, and the hypothesis that planetary formation is the cause is independently theoretically supported \citep{chambers2010}. The peculiar solar abundance therefore also contributes to the apparent overabundance of heavy elements in normal A-stars, and we may assume that a specific physical effect in A-star photospheres is not at play.
\item[3.] One must also consider that the abundances of heavier elements are derived from fewer lines, and are therefore at greater risk of having erroneous values caused by blends.
\end{enumerate}

Nevertheless, for those elements that were studied by \citeauthor{gebranetal2010} we also show the difference between KIC\,3429637 and normal A-star abundances, relative to the Sun in Table\,\ref{abundances}. Mn, Ni, Sr and Ba are strongly correlated with Fe \citep{gebranetal2008}, and Y and Zr increase rapidly with Fe/H \citep{gebranetal2010}, thus the chemical peculiarities of KIC\,3429637 would be even more pronounced were it not for its low iron abundance. Therefore in the final column of Table\,\ref{abundances} we tentatively provide the abundance differences as if KIC\,3429637 had the same Fe/H as the mean value of the `normal' A-stars to which we are comparing, namely [Fe/H] = 0.19. The correction is calculated from the data tables provided in the online material of \citet{gebranetal2010}. Since A-stars are known to show large star-to-star variations in abundances compared to F-stars \citep{gebranetal2008}, the correlation between [M/H] and [Fe/H] is not always tight, hence the `tentative' provision. Importantly, the elements for which the correlation is tighter are those that are the most peculiar. \citet{fossatietal2008} found that there is no temperature effect on abundance patterns, thus we need not worry about any temperature difference between KIC\,3429637 and the mean of the A-stars against which we are comparing.

We did not take non-LTE effects into account in our calculations, yet these effects can be important. Given the importance of calcium in Am star spectra, we considered non-LTE corrections for this element, but they are small \citep{mashonikaetal2007}. For iron, non-LTE corrections are significant ($0.2-0.3$\,dex) for \ion{Fe}{I}, but not \ion{Fe}{II}, and are positive \citep{rentzsch-holm1996}. The corrections would bring our Fe abundance closer to the solar value. Conversely, corrections for carbon are negative \citep{rentzsch-holm1996} and would further exaggerate the carbon deficiency seen in this star and expected in Am stars.

All three of our $T_{\rm eff}$, $\log g$ and $v\sin i$ determinations agree extremely well with the weighted means of those presented in \citet{catanzaroetal2011}, and $\xi$ is slightly high for a star of this temperature \citep{takedaetal2008}. We note, however, that in Am stars $\xi$ is expected to be slightly higher than in normal stars (\citealt{gebran&monier2007}; \citealt{coupry&burkhart1992}). The latter authors pointed out that the $\xi$ distribution reaches a maximum around A5, where the angular rotational velocity distribution also reaches a maximum. This could implicate rotational instabilities in generating microturbulence, but if that were true Am stars would not be observed to have higher $\xi$ values. \citet{paceetal2006} defined the relation $\xi = -4.7 \log(T_{\rm eff}) + 20.9$\,km\,s$^{-1}$ for horizontal branch stars, in which radiative levitation is also important -- this would give an expected $\xi$ of 2.7\,km\,s$^{-1}$, but the relation is not appropriate for the whole main sequence, and might even be used as a diagnostic to distinguish horizontal branch stars from main sequence stars.

The low $\log g$ value indicates this star is evolved. Interestingly, the value of $v\sin i$ is high for an evolved Am (i.e. $\rho$\,Pup) star -- Am Main-Sequence stars are already slow rotators, with the Am/Ap star unimodal rotation velocity distribution peaking at $\sim$60\,km\,s$^{-1}$ \textit{after} statistical correction for inclination \citep{abt&morrell1995}. During evolution, stars transfer angular momentum inward through radius expansion, slowing the surface rotation further. Hence by accepting the star is evolved, the high $v\sin i$ argues against the type of close binary system common among Am stars\footnote{Tidal braking is most significant when the orbital period decreases below 10\,d -- the tidal braking mechanism proposed by \citet{zahn1975} denotes that braking timescales depend on the stellar separation to the power 8.5. This is confirmed observationally by the fact that SB2 systems are circular for periods below $\sim$8\,d \citep{debernardi2000}.}. Instead, the hypothesis of a single A star with a sub-average rotation velocity is favoured for this case, in which this star was rotating slowly enough to develop some peculiarities.

It is well established that a large fraction of Am stars are found in binaries and that tidal braking slows the stellar rotation, allowing diffusion to occur. With the exclusion of Sc \citep{fossatietal2008a}, peculiarities are more extreme in more slowly rotating stars \citep{takedaetal2008}. The observed moderate $v\sin i$ value is indeed compatible with the fact that KIC\,3429637 is not an extreme Am star, as the moderate abundance anomalies imply. Notwithstanding, $\rho$\,Pup stars do not have extreme abundances, regardless of $v\sin i$. However, \citet{statevaetal2012} found that peculiarity is also slightly correlated with orbital eccentricity: more eccentric orbits produce more anomalous abundances, but we do not expect this to have any significance for wide binaries. Attempts to use the frequency modulation technique of \citet{shibahashi&kurtz2012} to infer orbital parameters were unsuccessful -- orbital sidelobes were not present. This indicates that either the star is in a binary system with an orbital period longer than the dataset, i.e. $P_{\rm orb} >$ 700\,d, or that the star is not binary at all.

\section{Astrophysical cause of amplitude growth}
\label{sec:cause}

The different amplitude growth rates of each mode, visible in Fig.\,\ref{fig:growth}, can be used to determine a time at which the mode amplitudes were too low to be detectable from the ground, assuming the growth rates are well behaved. A linear growth rate gave the best fit to $f_1$, but an exponential growth rate was more appropriate to $f_2$, which is growing much more rapidly than the lower frequency mode. Taking the limit of ground-based detection as 1\,mmag in the amplitude spectrum, $f_1$ would only have become detectable $\sim$2.6\,y ago, and $f_2$ $\sim$11\,y ago. We might conclude that these modes would have been undetectable in \citeyear{abt1984} at the time when \citeauthor{abt1984} spectroscopically observed the star to be metallic lined, although no photometric investigation into variability was carried out. However, \citet{breger2000b} observed the evolved $\delta$\,Sct star 4\,CVn over three decades and found mode amplitudes to be unpredictable and highly variable. One high-amplitude mode in that star disappeared altogether and re-emerged with a random phase. With so few studies of amplitude growth rates, and the unpredictable behaviour seen in 4\,Cvn, we cannot predict how KIC3429637 will behave or how it behaved in the past. This star is most certainly worthy of continued observation to monitor the amplitude changes. 

In any case, the current rates of growth indicate the star has either: (a) not been pulsating for very long, and has just crossed some instability threshold; (b) undergoes large amplitude variations over long periods of time, possibly in some cyclic nature; or (c) none of the above. We do not have enough data to pursue hypothesis (b), in that we have not observed a full cycle of variation in the light curve, but since the amplitude growth is not the same for each mode, nor does it appear to be sinusoidal, the cyclical amplitude variation hypothesis is unsupported by the present evidence. If we assume this star to be like 4\,CVn, then continued observations are required to analyse the changing amplitudes. In the meantime, let us examine hypothesis (a) from a theoretical point of view.

Even if the pulsations are not especially laminar, the destruction of the chemical abundance anomalies may take time, such as in the case of $o$\,Leo\,A, which is an evolved Am star whose abundance anomalies have not yet been erased \citep{michaudetal2005}. KIC\,3429637 could well have been a classical Am star on the main sequence that is now at or approaching a stage of rapid evolution in which its chemical peculiarities are being erased and its pulsation amplitudes are growing. Temporal evolution of abundance anomalies is another under-studied area of Am star research; KIC\,3429637 would make a perfect case-study. Existing studies (e.g. \citealt{gebranetal2010}; \citealt{abt1979}) have mostly focussed on multiple co-eval stars in one cluster and contrasting clusters of different ages. In these cases the scatter in abundances of given elements is large enough that no real difference of any element with age is prominent. Focussing on single, regularly-observed targets such as KIC\,3429637 is thus potentially astrophysically rewarding.

\section{Noise in the residuals}
\label{sec:noise}

\begin{figure*}
\begin{center}
\includegraphics[width=0.9\textwidth]{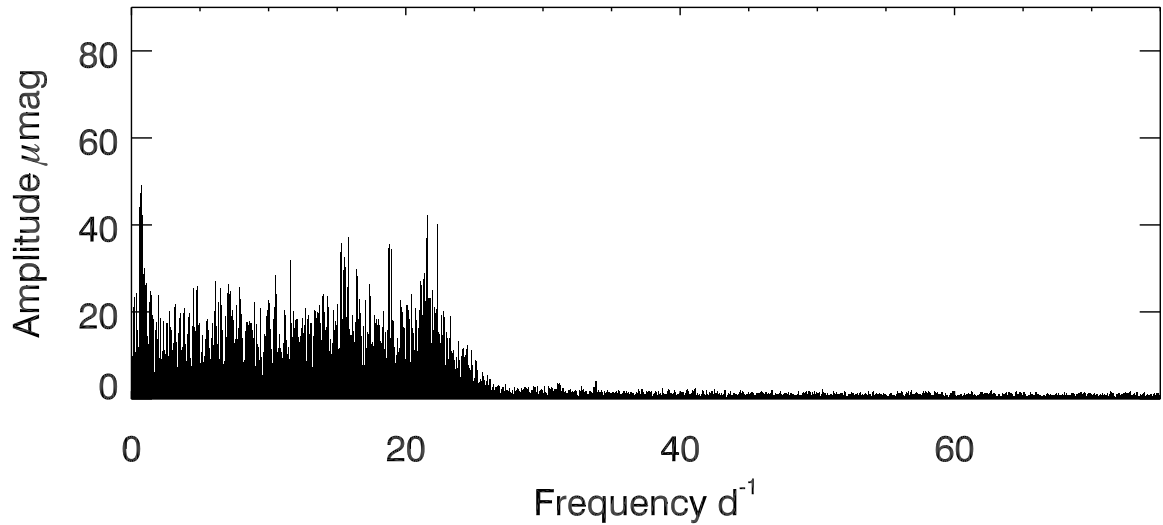}
\caption{There is a rapid decrease in noise level at around 24\,d$^{-1}$. Excess noise is injected by the PDC-LS pipeline at frequencies below this limit. There are no peaks remaining below that frequency with signal-to-noise $>$ 4.0, and no peaks have been removed at all above 24\,d$^{-1}$, hence the effect has not been caused by preferential prewhitening. Note that here we have prewhitened many more frequencies than were modelled in order to demonstrate the effect more clearly.}
\label{fig:granulation}
\end{center}
\end{figure*}

Spectroscopic data show KIC\,3429637 to be cool for a $\delta$\,Sct pulsator, and progressively cooler A-type stars have increasingly deep surface convection zones (see \citealt{kallinger&matthews2010} for a discussion). We used the Q7.1 SC data to look for such a signature of granulation in the form of a rapid change in noise with frequency. We present the results in Fig.\,\ref{fig:granulation}. After extensive investigation, the rapid decrease in noise seen between 20 and 30\,d$^{-1}$ was determined not to arise from granulation, but rather the noise is injected by the \textit{Kepler} PDC-LS\footnote{``LS'' describes the Least Squares algorithm used by the pipeline. We refer the reader to the papers of \citet{stumpeetal2012}, \citet{smithetal2012} and the Kepler Data Characteristics Handbook for more information on the pipeline, and to \citet{murphy2012} for its effect on asteroseismic analyses.} pipeline for this particular \textit{Kepler} quarter. It seems that approximately 15\,per\,cent of stars in the $\delta$\,Sct instability strip are afflicted with the same problem, which affects both long- and short-cadence data, but not necessarily all quarters for a given star. A more detailed description will be given in a future publication (Murphy, in preparation).

The Q8 data for this star contain no evidence for the ``granulation signature'', implicating noise injection by the pipeline over an astrophysical origin. The drop in noise at $\sim$24\,d$^{-1}$ in SC data is characteristic for this noise injection. The Nyquist frequency of LC data prevents the drop being seen in that cadence, but the noise is still present. The new PDC-MAP (Maximum A Posteriori) data will fix this for the LC data when the old data are reprocessed, but for now there is no implementation of MAP for SC.

\section{Structure and oscillation models}
\label{sec:models}

In our theoretical computations we used the Code Li{\'e}geois d'{\'E}volution stellaire CL{\'E}S \citep{scuflaireetal2008}, where the input physics is as follows: the equation of state is the superior `CEFF' (see, e.g. \citealt{christensen-dalsgaard&daeppen1992}); the opacity tables are OPAL opacities from \citet{iglesias&rogers1996} and \citet{alexander&ferguson1994} for high and low temperatures, respectively. The relative mixture of chemical composition from \citet{grevesse&noels1993} was used, while the convection was treated using Mixing-Length-Theory \citep[MLT]{bohm-vitense1958}. Our models include neither rotation nor diffusion. For our non-adiabatic, non-radial oscillation calculations we used the {\small MAD} code \citep{dupret2001}, where Gabriel's treatment of time-dependent convection has been implemented \citep{gabriel1996}.   

We consequently produced a grid, where $\ell = 0-3$ eigenfrequency spectra are computed for each of the models. For the selection of the best model we used the maximum likelihood method as described by \citet{grigahceneetal2012}.

\subsection{Results and discussion}

Using \textit{Kepler} data, \citet{uytterhoevenetal2011} showed how A-F pulsators can have such complex frequency spectra that they are challenging targets for any asteroseismic study. We consequently need as many constraints as possible for any theoretical modelling. Although the number of detected frequencies is reduced in KIC\,3429637 compared to what is usually detected in this type of star, its modelling is no less complicated.

\begin{figure}
\begin{center}
\includegraphics[width=0.49\textwidth]{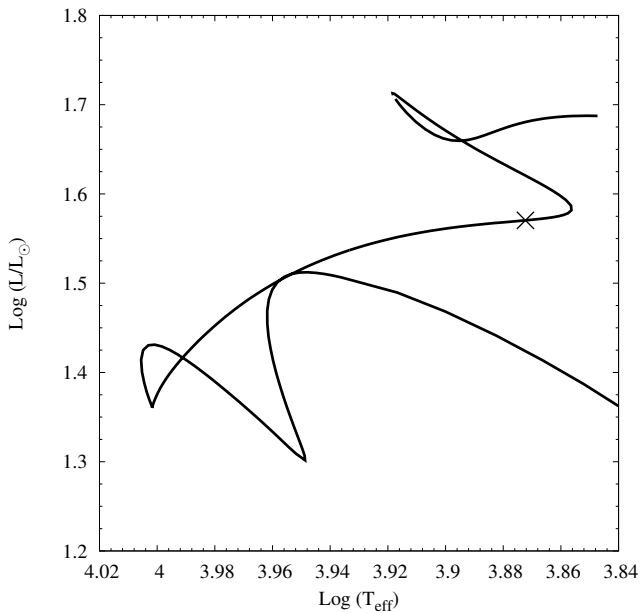}
\caption{Position of the best model on its evolutionary track in the HR diagram, showing the evolutionary stage of the star. As this is a model, there are no error bars.}
\label{fig:HR}
\end{center}
\end{figure}

We performed careful and detailed modelling, varying all the parameters (mass, metallicity, mixing-length parameter $\alpha_{\mathrm{MLT}}$, etc.) to obtain the combination giving the maximum value of the likelihood function. We explored different possibilities, taking into account the detected frequencies and their associated amplitudes, and the position of the star in the HR diagram. The main requirements on our models were: 1) fitting the maximum number of the observed frequencies, 2) modes should be unstable, and 3) the global parameters should be inside the observational photometric error box. In Table\,\ref{Param} we show the main properties of the best obtained model. All the parameters are within the spectroscopic uncertainties. We show the star's position on the HR diagram and its evolutionary stage in Fig.\,\ref{fig:HR}.

\begin{table*}
\centering
\begin{tabular}{c c c c c c c c c c }
M           &   T$_{\mathrm{eff}}$&   Log(L/L$_{\sun}$)  &   Log g &   R          &   Age   &   X     &   Z        &   $\alpha_{\mathrm{MLT}}$      &   $\alpha_{\mathrm{Ov}}$ \\ 
(M$_{\sun}$)  &   (K)           &                     &         &   (R$_{\sun}$)&   (y)    &        &           &         &    \\ 
\hline
 2.178 &   7452 &  1.570  &  3.648  &  3.666   &  8.982E+08 & 0.7360    & 0.0149   &2.0     &   0.2\\ 
\hline
\end{tabular}
\caption{Properties of the best model.}
\label{Param}
\end{table*}

Some important results come out from our comparison between theoretical time-dependent convection models and observations. First, we managed to obtain a model where seven of the eight highest-amplitude frequencies are excited. The lowest frequency mode (f$_{8}$=0.39448\,d$^{-1}$) is predicted to be stable (i.e. not excited). This mode might not be caused by oscillation -- at these low frequencies, instrumental origins are common (\citealt{murphy2012}, his Fig.\,3). The origin does not appear to be rotational or equal to the difference in frequency between any two other modes in Table\,\ref{Modes}. The alternative to instrumental origin is that it is indeed a pulsation frequency, in which case the conclusion is that some physics is missing from the model. We did not model all 40$+$ statistically significant frequencies due to the overwhelming complexity involved in doing so.

Second, we found that the value of the mixing-length parameter giving the maximum value for the likelihood function is $\alpha_{\mathrm{MLT}}$=2, i.e. different from the solar value in \citep{grigahceneetal2012}. It is known that $\alpha_{\mathrm{MLT}}$ should vary from star-to-star and also with temperature as the internal structure changes from a solar-like one to those of A- and F-stars. As this star is the first $\delta$\,Sct star to be treated with TDC models, coverage of $\alpha_{\mathrm{MLT}}$ for $T_{\rm eff} > 6000$\,K is currently unavailable, but for cooler stars we refer the reader to the works of \citet{grigahceneetal2012} and \citet{trampedach&stein2011} for deeper discussion.

A third remark regards the fitting of the frequency values. In Table\,\ref{Modes} we list the theoretical frequencies together with the observed frequencies and mode identifications. Four of the observed frequencies are very well fitted by the theoretical modes -- the difference is less than 0.03\,d$^{-1}$ (0.35 $\mu$Hz). For the other three the difference is less than 0.2\,d$^{-1}$. However, an ambiguity arises for some modes in that there are multiple mode identification possibilities of similar likelihood. This indicates that ground-based multi-colour photometric or spectroscopic observation campaigns are necessary to complete the mode identification. 

Finally, while the asteroseismic effective temperature agrees with the spectroscopic one to within less than 2\,$\sigma$, the $\log g$ value from the models (3.65) does not agree particularly well with that from spectroscopy ($3.0 \pm 0.1$). Similar model behaviour was seen in \citet{grigahceneetal2012}.

\begin{table}
\centering
\begin{tabular}{l c c r c r }
Id. &  f$_{\mathrm{Obs}}$  &    $\ell$  &  $n$  &    f$_{\mathrm{Theo}}$  &     f$_{\mathrm{Obs}}$ - f$_{\mathrm{Theo}}$ \\
 &  d$^{-1}$   &     &    &    d$^{-1}$   &     d$^{-1}$  \\
\hline

1 & 10.33759 & 0.0 &\,3.0 &  10.36443 &   -0.02684  \\
  &            & 1.0 &\,0.0 &  10.45143 &   -0.11384  \\
  &            & 3.0 & -3.0 &  10.22146 &  \,0.11613 \\
  &            & 2.0 & -1.0 &  10.50785 &   -0.17027  \\[1pt]

2 & 12.47161 & 0.0 &\,4.0 &  \phantom{11}12.4716063 &  \,\,\,2.0E-07  \\
  &            & 3.0 & -1.0 &  12.45554 &  \,0.01607 \\[1pt]
  
3 & 10.93641 & 3.0 & -2.0 &  10.93305 &  \,0.00336 \\
  &            & 1.0 &\,1.0 &  10.94367  &   -0.00725  \\[1pt]

4 & \phantom{1}9.71214 & 3.0 & -4.0 &  \phantom{1}9.71185 &  \,0.00029 \\
  &            & 2.0 & -2.0 &  \phantom{1}9.71928 &   -0.00714 \\[1pt]

5 & 12.65066 & 1.0 &\,2.0 &  12.82501 &   -0.17435  \\
  &            & 0.0 &\,4.0 &  12.47161 &  \,0.17905  \\
  &            & 3.0 & -1.0 &  12.45554 &  \,0.19512 \\
  &            & 2.0 &\,1.0 &  12.89238 &   -0.24172  \\[1pt]

6 & 13.50229 & 3.0 &\,0.0 &  13.67304 &  -0.17074 \\[1pt]

7 & 11.48287 & 2.0 &\,0.0 &  11.65746  & -0.17459 \\[1pt]
\hline
\end{tabular}
\caption{Identification of oscillating modes. Negative $n$ values denote mixed-modes with $g$-mode-like character. An additional observed frequency, f$_{8}$=0.39448\,d$^{-1}$, was modelled but was not found to be excited. Formal least-squares frequency precision is $3 \times 10^{-6}$\,d$^{-1}$ for $f_1$ and $f_2$, and $2 \times 10^{-5}$\,d$^{-1}$ for $f_3$. Since our models do not include rotation to lift the degeneracy, all modes presented are $m=0$.}
\label{Modes}
\end{table}

\subsection{Conclusions from modelling}
We have compared the observed frequencies of KIC\,3429637 to state-of-the-art $\delta$\,Sct models. The results of this comparison are that: 1) all the observed modes (those listed in Table\,\ref{Modes}) are predicted to be excited in our model except the lowest frequency, 2) four out of seven frequencies are fitted to within 0.26\,per\,cent, and 3) the varying amplitudes between quarters might be explained by mode interaction -- the modelled frequencies are all rather close and are perhaps subject to resonances as described in, e.g. \citet{dziembowski&krolikowska1985}. We identify many potential mixed modes, and it is noteworthy that $g$-modes that aren't even visible in the Fourier transform have the potential to affect $p$-mode frequencies \citep{buchleretal1997}.

\section{Conclusions}

The high-precision \textit{Kepler} data indicate two dominant modes of growing amplitude, with different growth rates, and a third mode of intermediate amplitude whose amplitude decreases. The mode amplitude growth appears to be intrinsic to the star.

Our spectroscopic data confirm literature classifications that this star is of the marginal Am type. The heavy elements are overabundant with respect to solar values, and the light elements are underabundant -- as one expects for Am stars. Notably, calcium is normal.

Mode amplitudes in $\delta$\,Sct stars can change by large amounts on short timescales, with changing growth rates (cf. 4\,CVn). The amplitude growth in this star appears much less chaotic than for 4\,CVn, and we have proposed evolution as the dominant cause, in that pulsation amplitudes are expected to increase naturally in evolving Am stars as the \ion{He}{II} convection zone deepens and picks up residual helium. We cannot definitively rule out that mode interaction is the cause, where energy is being transferred from one or more modes into those whose amplitudes are growing, except to say that we see no modes decreasing in amplitude by comparable amounts to those by which the dominant modes are growing. As such, this explanation seems viable only if one or more unseen modes -- perhaps $g$-modes or high-degree $p$-modes -- are transferring the energy.

For the first time, time-dependent convection has been used in models of a $\delta$\,Sct star observed with \textit{Kepler}. Despite the narrow observational constraints, the modelling was difficult. We determine a 2.18-M$_{\sun}$, 7450-K model with Z = 0.0149 to be the best solution (cf. Table\,\ref{Param}), with the star being 94\,per\,cent of the way through its main sequence lifetime (see Fig.\,\ref{fig:HR}), and we match four of seven mode frequencies to within 0.26\,per\,cent (Table\,\ref{Modes}). Multi-colour photometry and/or time-series spectroscopy would facilitate mode identification, and this target is most worthy of continual follow-up in \textit{Kepler}'s long-cadence mode, not least because it offers a chance to witness stellar evolution on human time-scales.

\section*{acknowledgements}

This research has made use of the SIMBAD database, operated at CDS, Strasbourg, France. Calculations have been partially carried out in Wroclaw Centre for Networking and Supercomputing (http://www.wcss.wroc.pl), grant No. 214. Based on data from the Kepler Space Telescope, for which we are extremely grateful, and on observations made with the Nordic Optical Telescope, operated on the island of La Palma jointly by Denmark, Finland, Iceland, Norway, and Sweden, in the Spanish Observatorio del Roque de los Muchachos of the Instituto de Astrof\'{\i}sica de Canarias (IAC).

SJM would like to acknowledge the financial support of the STFC. AG acknowledges the KITP staff of UCSB for their warm hospitality during the research programme ``Asteroseismology in the Space Age". This research was supported in part by the National Science Foundation of the United States under Grant No.\ NSF PHY05--51164. EN acknowledges support from Polish MNiSW5 grant N N203 405139. KU acknowledges financial support by the Spanish National Plan of R\&D for 2010, project AYA2010-17803. We thank the Spanish Night-Time Allocation Committee (CAT) for awarding time to the proposal 61-NOT7/10A.

\bibliography{kic3429637_arXiv}

\end{document}